# Predicted Oxidation of CO Catalyzed by Au Nanoclusters on a Thin Defect-Free MgO Film Supported on a Mo(100) Surface

Chun Zhang, Bokwon Yoon, and Uzi Landman*

*School of Physics, Georgia Institute of Technology, Atlanta, GA 30332-0430*

**RECEIVED DATE (automatically inserted by publisher)**; uzi.landman@physics.gatech.edu

Recently there has been a surge in research pertaining to the physical and chemical properties of gold nanoclusters. Unlike supported particles of larger sizes, or extended solid surfaces[1,2], small metal clusters adsorbed on support materials were found to exhibit unique properties that originate from the highly reduced dimensions of the individual metal aggregates[3-8]. In particular we note here joint experimental studies and theoretical investigations[5-8] using first-principles simulations on size-selected small gold clusters, $Au_n$ (20 ≥ n ≥ 2), adsorbed on a well characterized metal-oxide support under ultra-high-vacuum conditions. These studies revealed that cluster morphology, dimensionality, dynamical structural fluxionality, electronic structure, and charge state, as well as the state of the metal-oxide surface (specifically, defect-rich MgO(100) surfaces containing oxygen-vacancy $F$ centers, or defect-poor surfaces), govern the catalytic activity. Particular emphasis has been put[5-8] on the low-temperature (as low as 140K) oxidation of CO, catalyzed by $Au_n$ clusters with 7< n < 25 atoms that are adsorbed on F-center-rich magnesia surfaces. In these investigations charging of the adsorbed metal cluster through partial electron transfer from the substrate F-center defect and subsequent occupation of the antibonding $2\pi^*$ orbital of $O_2$ leading to activation of the O-O bond of the molecule adsorbed on the cluster (resulting in formation of peroxo or superoxo states), have been identified as underlying the catalytic activity.

Here we show results of first-principles investigations aiming at tuning and controlling the catalytic activity of gold nanoclusters through the design of the underlying support. We show that gold clusters adsorbed on a very thin (2 layers) defect-free MgO film which is itself supported on Mo(100)[9], may serve as model catalysts for the low-barrier oxidation of CO. The origin of the emergent activity of the nanoclusters is a dimensionality crossover of the adsorbed gold clusters, from inactive three-dimensional (3D) optimal structures on thick MgO films, to catalytically active 2D ones for sufficiently thin MgO films (less then 1nm in thickness) supported on Mo(100). The increased gold wetting propensity on the MgO/Mo(100) surface originates from electrostatic interaction between the underlying metal and metal-induced excess electronic charge accumulated at the cluster interface with the metal-oxide film[10]. The excess interfacial charge is predicted here to activate $O_2$ molecules adsorbed at the interfacial periphery of the 2D gold island with the MgO/Mo(100) surface. This activation, which weakens greatly the O-O bond, lowers rather remarkably the barrier for reaction of the activated molecule with CO and the subsequent emission of $CO_2$.

Our first-principles calculations are based on a density functional theory approach[11,12] with exchange and correlation energy corrections included through a generalized gradient approximation (GGA)[13]. A plane wave basis is used with a cutoff kinetic energy of 30 Ry, and ultrasoft pseudopotentials (scalar relativistic for Au)[14] are employed with Γ–point sampling of the Brillouin zone. In structural relaxations corresponding to minimization of the total energy, convergence is achieved when the forces on the atoms are less than 0.001 eV/Å.

In modeling the metal-supported MgO films we use a four-layer Mo(100) slab (lattice constant of 3.15 Å) of thickness 4.64 Å which has been found to reproduce (in its middle) the bulk electronic properties of Mo[15]. We adsorb the planar $Au_{20}$ clusters on a MgO/Mo(100) with 5 x 6 unit cells of the Mo(100) surface; for the 3D.tetrahedral cluster the dimensions of the MgO surface (without the metal support) are the same, and the bottom layer of the MO oxide is held fixed. In all calculations, the periodically replicated slabs are separated [in the (100) direction] by a vacuum region of 20 Å. In structural optimizations all the atoms of the adsorbed gold clusters, the MgO thin film, and the first two layers of the Mo substrate are allowed to relax. In calculation involving the tetrahedral gold structure the MgO(100) crystalline surface was modeled by a two-layer MgO(100) slab which is sufficiently thick to both reproduce the properties of the bare MgO surface[16] and to obtain converged results (with respect to the number of MgO layers) for the energetics of adsorbed Au clusters[17]. For further details see ref. 10

The gas-phase optimal 3D tetrahedral $Au_{20}$ cluster[18] maintains it's structure on the MgO(100) surface, with a 1.2 eV advantage over the planar structure. However, this cluster is found to adsorb $O_2$ only weakly (0.34 eV on the top apex atom of the pyramid with (d(O-O)=1.28 Å remaining close to the gas phase value), and no binding is found at peripheral sites of the gold/MgO interface.; when CO is preadsorbed to the top apex Au atom (0.7 eV) no coadsorption of $O_2$ occurs.

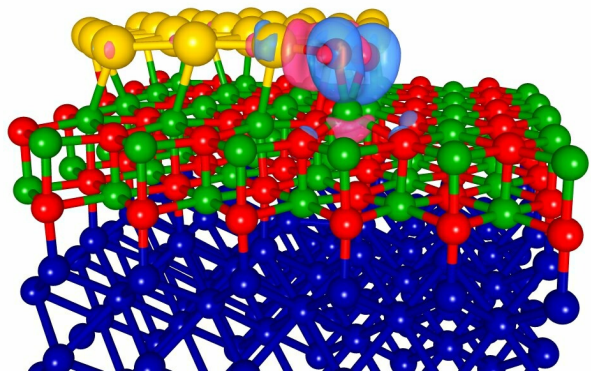

*Figure1*. Two dimensional $Au_{20}$ island (yellow on line) adsorbs on a 2-layer MgO film (O atoms in red and Mg in green) supported on Mo (100) (blue on line), with a coadsorbed $O_2$ molecule. Superimposed we show an isosurface of the excess electronic charge (light blue on line) illustrating activation of the adsorbed molecule through population of the antibonding $2\pi^*$ orbital.

In light of the above inactivity of the 3D structure we focused our investigations on $Au_{20}$ adsorbed on a two-layer MgO/Mo(100). Here the planar isomer is more stable then the 3D one by 3.3 eV, with the enhanced stability resulting from penetration of metal states through the thin MO film and charge accumulation at the cluster/MgO interface (up to 1e for $Au_{20}$)[10]. Furthermore, in this configuration the $O_2$ adsorbs relatively strongly (1.35 eV) at the cluster periphery (Fig. 1), and the process is accompanied by transfer of electronic charge [19] (about 1.3e) into the antibonding $2\pi^*$ orbital leading to activation of the O-O bond of the adsorbed molecule into a peroxo state with d(O-O) = 1.52 Å and no spin polarization.

We explored two mechanism for the reaction of CO with the activated oxygen molecule. One involves a coadsorbed CO molecule (Langmuir-Hinshelwood, LH) and in the other one the reactant CO molecules approaches the activated $O_2$ directly from the gas phase.(Eley-Rideal, ER).Various CO coadsorption sites were explored. First we examined adsorption to the gold cluster, yielding relaxed adsorption configurations with CO binding energies ranging from 0.8 eV (CO binding to a peripheral Au atom that is nearest-neighbor to the Au atom bonded to the $O_2$), down to a vanishingly small binding for a non-periphery nearest-neighbor Au atom on the flat gold island. However, calculation of the barrier[20] for reaction between the reactants (with CO at the peripheral site) yielded a high value of close to 1.0 eV.

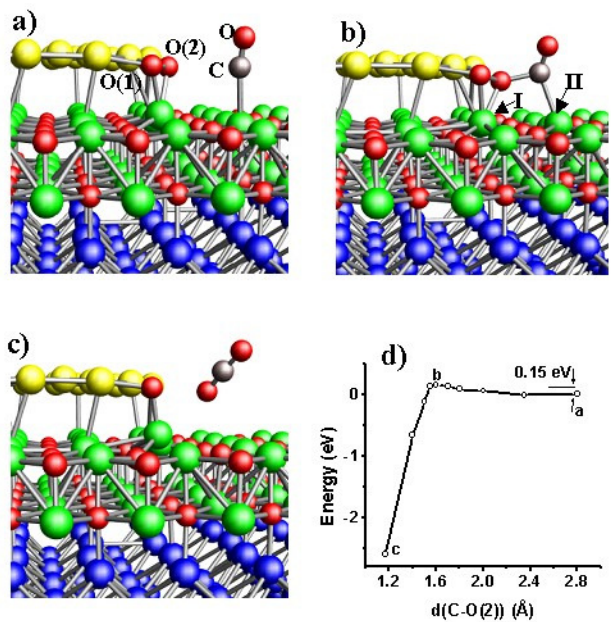

*Figure 2.* (a-c) Configurations of the two dimensional $Au_{20}$ island shown in Fig. 1 (the color scheme is the same as in Fig.1) with coadsorbed $O_2$ (the atoms marked O(1) and O(2), and CO (C atom in gray online). (a) the initial optimized configuration; d(O(1)-O(2))=1.52Å, d(C-O(2))=2.85Å. (b) The transition state (the nearest neighbor Mg atoms are marked as I and II. Distances are: d(O(1)-O(2))=1.55Å, d(C-O(2))=1.60Å, d(C-O)=1.18Å, d(C-Mg(II))=2.30Å..(c) a configuration illustrating formation and desorption of $CO_2$ . (d) The total energy profile along the C-O(2) reaction coordinate, with the zero of the energy scale taken for configuration (a). The sharp drop past the barrier top corresponds to $CO_2$ formation.

We have tested also several CO coadsorption sites located on the MgO surface in the vicinity of the adsorbed $O_2$. The adsorption geometry shown in Fig. 2(a) is characterized by a CO binding energy of 0.4 eV. Starting from this coadsorption configuration the LH reaction barrier leading to formation of $CO_2$ was calculated to be 0.15 eV; the transition state is shown in Fig.2(b), $CO_2$ formation in Fig. 2(c), and the energy profile along the reaction coordinate is displayed in Fig. 2(d). The product molecule desorbs readily in the course of the reaction (the binding energy of $CO_2$ to a Mg site on the MgO(100) surface is merely 0.14 eV). Alternatively, we considered several reaction trajectories that follow the ER mechanism. Some of these trajectories yielded formation of $CO_2$ with a small (0.2 eV) or vanishngly small barriers, depending on the angle of approach.

In summary, with the use of first-principles simulations we predict low barrier CO oxidation reactions to occur via a Langmuir-Hinshelwood or an Eley-Rideal mechanism, on 2D gold nanocluster islands adsorbed on very thin (2 layers) MgO films that are supported on Mo(100). Underlying the catalytic activity, that is predicted to occur even in the absence of F-center defects, is the excess electronic charge at the gold cluster/magnesia interface. This results from penetration of metal states through the MO thin film. We expect these results to provide the impetus for future experiments, as well as the further development of methods for controlling the interfacial charge (and consequently the chemical reactivity) in nanocatalytic systems, for example, through the use of applied fields.

**Acknowledgement**: This research was supported by the US AFOSR and the DOE. Computations were performed at the DOE National Energy Research Scientific Computing Center, at the Lawrence Berkeley National Laboratory (NERSC) and at the Georgia Tech Center for Computational Materials Science.

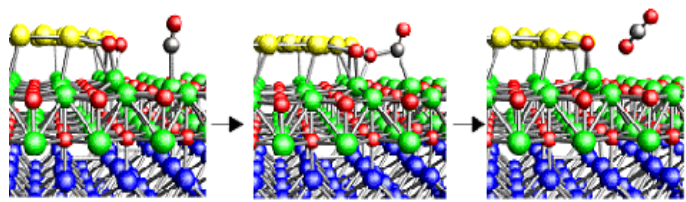

Configurations of the two-imensional $Au_{20}$ with coadsorbed $O_2$ and CO. Color scheme: gold atoms in yellow, carbon in gray,, oxygen in red, Mg in green, and Mo in blue. From left to right: the initial optimized configuration; the transition state; a configuration illustrating formation and desorption of $CO_2$. In the rightmost panel he total energy profile along the C-O(2) reaction coordinate is displayed, with the zero of the energy scale taken for the initial configuration. The sharp drop past the barrier top corresponds to $CO_2$ formation